%% file: main.tex
\documentclass[conference]{IEEEtran}
\IEEEoverridecommandlockouts
\usepackage[hyphens]{url}
\usepackage{hyperref}
\usepackage[hyphenbreaks]{breakurl}
\usepackage{cite}
\usepackage{tcolorbox}
\usepackage{paralist}
\usepackage{graphicx}
\usepackage{xspace}
\usepackage{tabu}
\usepackage{makecell}
\usepackage{booktabs,multicol,multirow,graphicx,tabularx,ragged2e,dcolumn}
\usepackage{enumitem}

\makeatletter
\newcommand{\etal}{\emph{et al}.\@\xspace}
\newcommand*{\eg}{\emph{e.g.},\@\xspace}
\newcommand*{\ie}{\emph{i.e.},\@\xspace}
\newcommand*{\aka}{\emph{a.k.a.}\@\xspace}
\def\BibTeX{{\rm B\kern-.05em{\sc i\kern-.025em b}\kern-.08em
    T\kern-.1667em\lower.7ex\hbox{E}\kern-.125emX}}
\begin{document}
\newcommand{\linebreakand}{%
  \end{@IEEEauthorhalign}
  \hfill\mbox{}\par
  \mbox{}\hfill\begin{@IEEEauthorhalign}
}
\makeatother
\title{A Qualitative Study on the Sources, Impacts, and Mitigation Strategies of Flaky Tests}

\author{
\IEEEauthorblockN{Sarra Habchi}
\IEEEauthorblockA{
University of Luxembourg \\
sarra.habchi@uni.lu}
\and
\IEEEauthorblockN{Guillaume Haben}
\IEEEauthorblockA{
University of Luxembourg \\
guillaume.haben@uni.lu}
\and
\IEEEauthorblockN{Mike Papadakis}
\IEEEauthorblockA{
University of Luxembourg \\
michail.papadakis@uni.lu}
\linebreakand
\IEEEauthorblockN{Maxime Cordy}
\IEEEauthorblockA{
University of Luxembourg \\
maxime.cordy@uni.lu}
\and
\IEEEauthorblockN{Yves Le Traon}
\IEEEauthorblockA{
University of Luxembourg \\
yves.letraon@uni.lu}
}
\maketitle

\begin{abstract}
   Test flakiness forms a major testing concern. Flaky tests manifest non-deterministic outcomes that cripple continuous integration and lead developers to investigate false alerts.
   Industrial reports indicate that on a large scale, the accrual of flaky tests breaks the trust in test suites and entails significant computational cost. 
   To alleviate this, practitioners are constrained to identify flaky tests and investigate their impact.
   To shed light on such mitigation mechanisms, we interview 14 practitioners with the aim to identify (i) the sources of flakiness within the testing ecosystem, (ii) the impacts of flakiness, (iii) the measures adopted by practitioners when addressing flakiness, and (iv) the automation opportunities for these measures.
   Our analysis shows that, besides the tests and code, flakiness stems from interactions between the system components, the testing infrastructure, and external factors.
   We also highlight the impact of flakiness on testing practices and product quality and show that the adoption of guidelines together with a stable infrastructure are key measures in mitigating the problem. 
   %   both monitoring and static analysis tools are needed to assist these measures.

\end{abstract}


%\keyword{flaky tests, integration testing, empirical study}

\maketitle

\input{introduction}
\input{related_works}
\input{GLR}

\input{study_design}
\input{study_results}
\input{threats}

\input{conclusion}

\bibliographystyle{IEEEtran}
\bibliography{Flakiness}
\end{document}

%% file: introduction.tex
\section{Introduction}
\label{intro}
%Integration testing is the process of combining software units that were developed separately to test if they work together~\cite{Integrat56:online}.
Software Testing is critical for modern software development as it allows the concurrent implementation and integration of features. %,  integrate well and do not break existing ones.
At Google, more than 50 million test cases are executed every day to ensure the quality of their products~\cite{Welcomet41:online}.
%How important is integration testing: costs spent on it
%However, in the last few years,
Though, test automation faces major problems with the emergence of  flaky tests % as one of the main threats to the reliability of this process
~\cite{FlakinessGoogle,Mozilla,FlakinessSpotify}. Flaky tests are tests that, for the same versions of code and test, can pass and fail on different runs.
Such non-determinism sends confusing signals to developers who struggle to interpret the test results. %  and need to decide whether or not to integrate the new features
As a result, developers lose trust in test suites, disregard their signals and integrate features containing real failures, thereby nullifying the purpose of testing.

Flaky tests are prevalent in large software systems and they incur significant cost.
Google reports indicate that 16\% of their tests exhibit some flakiness whereas 84\% of the transitions from pass to fail involve a flaky test~\cite{FlakinessGoogle}.
This entails enormous computational resources since 2-16\% of the company's testing budget is dedicated to rerun flaky tests~\cite{GTAC2016}.
%or reruns and introduce costly delays into the core development process~\cite{FlakinessGoogle}.
%Add other examples from the Apple paper

In response to this challenge, researchers dedicate their efforts in understanding the nature of flaky tests and the way they manifest.
Empirical studies examined the root causes of flaky tests in open-source software~\cite{luo_empirical_2014,eck_understanding_2019,Thorve2018,dutta_detecting_2020} and industrial systems~\cite{Lam2019RootCausing}, showing that concurrency and order-dependency are among the main categories of test flakiness. 
Notably, the study of Eck~\etal~\cite{eck_understanding_2019} showed that flakiness can stem from the code under test and highlighted its potential impact on organisational aspects like resource allocation.

Other studies investigated tools and techniques that could help developers to cope with test flakiness.
Automated tools, such as DeFlaker~\cite{bell_deflaker_2018}, iDFlakies~\cite{lam_idflakies_2019}, and FlakeFlagger~\cite{alshammari2021flakeflagger} have been developed in order to detect flaky tests with a minimum number of test runs or re-runs. Unfortunately, these advances offer only partial solutions to the problem and may not fit well within the development systems and organisation constraints. For instance, DeFlaker relies on coverage and reruns of tests that do not execute changed code, which are not possible in specific development environments that use regression test selection or when coverage cannot be obtained.     
Furthermore, the fixing of flaky tests gained traction as studies investigated the fixing effort devoted for flaky tests and tools like \cite{shi_ifixflakies_2019} are designed to fix flaky tests.
Nonetheless, in order to devise flakiness solutions, we need to understand how developers deal with flaky tests in practice.
In particular, it is necessary to identify the typical measures taken by practitioners when dealing with flaky tests, and reflect on how research solutions could assist and improve them.
% Hence, in order to devise solutions, we need to understand how developers deal with flaky tests in their daily work. In particular, it is necessary to identify the typical measures taken by practitioners when dealing with flaky tests, and reflect on how research could assist and improve them.

To shed some light on these questions, we conduct an empirical study focused on the industrial context in which flakiness manifests. % and the current state of practice regarding its mitigation.
Specifically, we perform a qualitative analysis on data collected from 14 practitioner interviews to answer the following research questions:

\begin{itemize}[wide=10pt,noitemsep,topsep=0pt]
    \item \textbf{\textsc{Rq1:}} Where can we locate flakiness?\\
    \textit{\textbf{Goal:}} Differently from previous studies~\cite{luo_empirical_2014,eck_understanding_2019,Thorve2018,dutta_detecting_2020,Lam2019RootCausing}, which focused on the root causes of flakiness, \eg concurrency and timeouts, we aim to identify where flakiness stems within the different components of the development ecosystem, \eg test, code under test, and infrastructure. This localisation is necessary to guide both detection and fixing approaches.\\
    \textit{\textbf{Results:}} In addition to tests, flakiness stems from the poor orchestration between the system components, the testing infrastructure, and external factors, \eg OS and firmware. Studies should consider and leverage these factors when addressing flaky tests and not focus solely on the test and code under test.
    
    \item \textbf{\textsc{Rq2:}} How do practitioners perceive the impact of flakiness?\\
    \textit{\textbf{Goal:}} This question is commonly discussed in industrial reports and research studies. In this paper, 
    we examine it through direct discussions with practitioners. The aim is to understand the impact of flakiness on the development workflow and practices. \\
    \textit{\textbf{Results:}} Besides dissipating development time and hindering the continuous integration (CI), flakiness affects the testing practices and leads to a degradation of the system quality. We also shed light on the pernicious consequences of system flakiness, \ie buggy or non-deterministic features that are falsely labelled as flaky tests.
    
    \item \textbf{\textsc{Rq3:}} How do practitioners address flaky tests?\\
    \textit{\textbf{Goal:}} This question aims at identifying and understanding the measures taken by practitioners to address flakiness before and after it manifests in the CI.\\
    \textit{\textbf{Results:}} 
    %Our analysis shows that depending on their resources, practitioners deal with flakiness differently. 
    The prevention of test flakiness is performed by building stable infrastructures and enforcing guidelines, whereas the detection still relies mainly on reruns and manual inspection. Our results also highlight monitoring and logging tasks, which are commonly dismissed in research, yet they are key to most of the mitigation measures taken by practitioners.
    
    \item \textbf{\textsc{Rq4:}} How could mitigation measures be improved with automation tools?\\
    \textit{\textbf{Goal:}} This question aims to identify specific needs to be addressed by future research. \\
    \textit{\textbf{Results:}} We accentuate the need for techniques that monitor and analyse the system states to assist the prediction, debugging, and fine-grained evaluation of flaky tests.
    Our participants also expressed the need for automating the quality assessment of software tests through static analysis and variability-aware reruns, \ie reruns under diverse system configurations. 
    %confirm previous findings about the difficulty of reproducing and identifying the root causes of flaky tests~\cite{eck_understanding_2019}
\end{itemize}

We believe that the qualitative results of this study are necessary to complement the current understanding of flakiness and advise future work.  
%Therefore, we discuss at the end of this paper a list of actionable implications of our study.

%The remainder of this paper is organised as follows.
%Section~\ref{sec:glr} presents our process for collecting data from grey literature, while Section~\ref{sec:interviews} describes the interviewing process and the analysis of collected data. Section~\ref{sec:results} reports and discusses the results and Section~\ref{sec:threats} discusses the potential threats to the study validity.
%Finally, Section~\ref{sec:related} presents the relevant literature and Section~\ref{sec:conclusion} concludes with a discussion of the study implications.

%% file: related_works.tex
\section{Related Work}
\label{sec:related}

The first study on test flakiness was carried out by Luo \etal~\cite{luo_empirical_2014}. 
They analysed 201 commits from 51 open source projects in order to understand the root causes of flaky tests. % and the strategies used by developers to fix them. 
They showed that Async Waits, concurrency, and test order-dependency are the main categories of flakiness. 
Later Lam \etal~\cite{Lam2020} analysed 55 Java projects to understand the introduction of flakiness in software systems. 
They found that 75\% of the 245 detected flaky tests were already flaky when added to the test suite, thus justifying the need to run detectors on newly-introduced tests. 
Flakiness has also been studied in specific software systems. 
%Two studies \cite{Cordy2019,Qin2021} stressed the importance of accounting for flakiness in research by highlighting the impact that it can have on mutation testing and automated program repair.
For instance, Luo \etal was replicated by Thorve \etal~\cite{Thorve2018} on Android applications where they found new root causes of flakiness: dependency, program Logic, and UI. 
Dutta \etal~\cite{dutta_detecting_2020} conducted an extensive study on flaky tests (causes and fixes) in machine learning applications, finding that algorithmic non-determinism represents the biggest source of flakiness.

%Several studies have been conducted to inspect flakiness in industrial contexts.
%highlighting the importance of this the challenge it represents and the necessity for solutions to this problem.
Lam \etal conducted two studies \cite{Lam2019RootCausing,lam_study_2020} about flaky tests at Microsoft. The first study showed that the number of build failures can quickly become significant despite having a low number of flaky tests. 
Thence, they introduced \textit{RootFinder}, a tool that identifies the root causes of flaky tests by analysing differences in test logs and spotting suspect method calls. 
In their second study, Lam \etal presented \textit{FaTB}, an automated tool that speeds the runtime of test suites by lowering timeouts and waits without impacting the overall test suite flake rate. 
Leong \etal~\cite{LeongSPTM19} studied flaky tests at Google and found that more than 80\% of test output transitions are caused by flakiness. 
% In their study, they also showed that flakiness systematically overestimated the performance when evaluating regression testing techniques.
At Apple, Kowalczyk \etal~\cite{Kowalczyk2020} introduced a flakiness scoring system and showed its ability to reduce flakiness by 44\%. 

Eck \etal~\cite{eck_understanding_2019} surveyed 21 Mozilla developers, asking them to classify 200 flaky tests in terms of root causes and fixing efforts. 
%They found that flakiness is perceived as a significant problem regardless of the team and project size. 
This study highlighted four new categories of flakiness: restrictive ranges, test case timeout, test suite timeout, and platform dependency.
It also provided evidence about flakiness from the CUT and showed that flaky tests can have organisational impacts.
In this study, we leverage a different qualitative approach (interviews) to address other aspects of flakiness in practice.
More specifically, we investigate broader sources of flakiness (\eg SUT and infrastructure) instead of the root causes (\eg concurrency and timeouts).
Our study also inspects the actions taken by practitioners in order to prevent, detect, and alleviate flaky tests.
Result-wise, our findings confirm the observations of Eck \etal about (i) the impact of flakiness on the test suite reliability and (ii) the challenges of reproducing and debugging flaky tests.
Furthermore, we highlight new flakiness impacts, on testing practices and product quality, and we synthesise a list of automation challenges for flakiness mitigation.

%Finally, \etal~\cite{ahmad_empirical_2019} carried out another survey of developers from 4 Swedish companies and identified 23 factors affecting flakiness.

%Add result wise

%% file: GLR.tex
\section{Preliminary analysis: Grey Literature Review}
\label{sec:glr}
%The main objective of this study is to understand the measures adopted by practitioners when dealing with flaky tests.
We conduct a grey literature review (GLR) to establish an initial mapping of the measures adopted by practitioners when dealing with flaky tests.
This mapping lays the foundation for our mitigation analysis (RQ3) and helps in guiding our interview design.
With respect to this objective, this GLR is exploratory and non-exhaustive.  
In the following, we explain our process for collecting, evaluating, and analysing data from the grey literature.
\paragraph{Search}
We followed the recommendations of Kitchenham and Charters~\cite{kit_cha_2007}, for the reviewing process in general, and the guidelines of Garousi \etal~\cite{garousi2019guidelines} for the aspects specific to grey literature.
The research question for our review is:
\begin{compactitem}
    \item \textsc{\textbf{RQ3:}} How do practitioners address flaky tests?
\end{compactitem}
In order to answer this question, we focused our review on materials published by practitioners describing their mitigation of flakiness, \eg technical reports, presentations, blogs, etc.
To collect these materials, we queried the advanced Google search engine with the following string:
\texttt{(Mitigate OR Manage OR Deal OR Control OR Avoid OR Prevent OR Tools OR Identify OR Detect) AND (Flaky OR Intermittent OR Unreliable OR non-deterministic) AND Tests}.
This query resulted in $276,000$ results.
We manually checked the top 100 articles and only accepted articles that:
\begin{itemize}
    \item Are written by practitioners. Articles and Blog posts written by researchers are excluded.
    \item Depict practitioners' views on flakiness and do not only address the problem theoretically.
\end{itemize}
We found that only 56 articles correspond to the searched material as a large part of the top-100 articles were dedicated to the introduction of flakiness without addressing its mitigation.
% Indeed, many articles are not based on practitioner views and they only explain the issue of flaky tests theoretically.
% We also found several formal literature articles or blogs that are inspired by them.
% We excluded these articles and focused on sources that reflect practitioner practices.
%The search engine and query 
%The inclusion criteria
\paragraph{Analysis}
The objective of this step is to identify and categorise the flakiness mitigation measures from the selected articles.
For this purpose, we first examined the 56 articles to check their adequacy for our analysis.
We relied on the quality assessment checklist presented by Garousi \etal~\cite{garousi2019guidelines}, which is specifically designed for grey literature sources.
We found that three factors are particularly relevant in our context and we adopted them as exclusion criteria:
\begin{itemize}[wide=10pt,noitemsep,topsep=0pt]
    \item \textbf{Objectivity:} We exclude sources where the authors have a clear vested interest. For instance, articles that promote new tools or plugins for mitigating flaky tests are generally biased.
    \item \textbf{Method adequacy:} We found that very few sources have a clearly stated their aim and methodology. However, from the presented content, we could identify articles that were not based on practical experience and exclude them. 
    For instance, in several cases, the authors present mitigation measures from a compilation of other sources and not based on their own experiences.
    \item \textbf{Topic adequacy:} We checked whether the articles enrich our analysis or not. More specifically, we excluded articles that do not present any mitigation measures for flaky tests.
\end{itemize}
The full quality assessment is available with our artefacts~\cite{artefacts}.
Based on the three exclusion criteria, we selected 38 articles that fit within the study scope and objectives.
Two authors read these articles and iteratively synthesised a classification of the measures described by practitioners.
This consensual process is similar to the qualitative analysis performed on the interview transcripts (\textit{cf.} Section~\ref{sec:interviews}).
%We found that the adopted strategies fall in three major categories: prevention of flaky tests, detection of flakiness before or after its manifestation, and mitigation actions taken when a test is identified as flaky.
The results of this analysis are presented in Table~\ref{table:strategies} and will be discussed in Section~\ref{sec:results}.
Interestingly, in our grey literature analysis, we observed that the articles do not explain the rationale behind the choice of measures.
Similarly, the consequences of the measures are generally dismissed.
Hence, we try to address these gaps in our interviewing process.

%% file: study_design.tex
\section{Interviews \& Analysis}
\label{sec:interviews}

% Our initial goal is to explore with an open-minded the following topics:
% \begin{itemize}
%     \item The sources of flaky tests;
%     \item The strategies for mitigating flaky tests;
%     \item The automation needs for flakiness mitigation; 
% \end{itemize}
The objective of the interviewing process is to explore the topics of our research questions with an open mind instead of testing pre-designed questions.
For this purpose, we pursue a qualitative research approach~\cite{creswell2017research} based on classic Grounded Theory concepts~\cite{adolph2011using}. 
In this section, we explain our implementation of this approach from the interview design to the analysis of the results.
\subsection{Questions}
Since we already formulated our topics of interest (RQs), we opted for semi-structured interviews.
These interviews build on starter questions, which cover the topics of interest, and according to the interviewee's answers, they develop follow-up questions that explore other points.
While designing and conducting our interviews, we followed the recommendations of Hove~\etal~\cite{hove2005experiences}.
In particular, we ensured the clarity of the discussed topics and notions before going through the interviews.
For instance, we always asked questions about the interviewee's definition of flakiness to avoid misunderstandings and ensure that the following questions are interpreted correctly.
We also avoided making prior assumptions about participants' opinions or actions.
For example, we ask several questions about the testing practices before formulating our questions to avoid wrong assumptions about the use of automated testing or CI. 
We also explained the non-judgemental nature of the interviews and encouraged participants to express their opinions freely.
Specifically, we mentioned that the objective is not to assess the participants' knowledge about flakiness but rather to grasp their perception of it.
Finally, we asked follow-up questions whenever possible and we favoured open-questions such as \textit{``Why did you opt for this measure?''} to incite participants to explain their motivations.
We structured our interview around the three following sections.
\paragraph{\textbf{Context}}
We asked questions to characterise the project and testing infrastructure.
\begin{enumerate}[leftmargin=*,noitemsep,topsep=0pt]
    \item What kind of projects do you work on? If possible ask for metrics like codebase size, architecture, and development team size.
    \item Do you have automated or manual tests?
    \item What kind of tests do you generally write?
    \item Do you have a continuous integration?
    \item Do you have a testing policy?
    \item Can you describe your testing infrastructure? Do you consider it stable?
\end{enumerate}

\paragraph{\textbf{Flakiness}}
We asked general questions about flakiness:
\begin{enumerate}[leftmargin=*,noitemsep,topsep=0pt] \setcounter{enumi}{6}
    \item Do you know what a flaky test is? 
    \item What is your definition of flakiness?
    \item How commonly do you encounter flaky tests?
    \item What are the sources of flakiness in your context? 
    \item Do you consider flakiness as an issue? Why?
\end{enumerate}

\paragraph{\textbf{Measures}}
We asked questions about the actions taken by participants to prevent and address flaky tests:
\begin{enumerate}[leftmargin=*,noitemsep,topsep=0pt] \setcounter{enumi}{11}
    \item How do flaky tests manifest in your codebase? How do you detect them?
    \item How do you treat the identified flaky tests?
    \item Do you adopt any specific measures to avoid flaky tests?
    \item Why did you adopt these measures?
    \item Do you face difficulties when dealing with flaky tests? 
    \item If yes: What are these difficulties and what could help you to overcome them?
\end{enumerate}
For each measure described by the participant, we asked follow-up questions to understand the motivations and consequences.
When possible, we also asked follow-up questions about the measures that the participants did not take, \eg if they never mention fixing flaky tests, we could ask about the rationale behind it.
%More specifically, when developers do not mention some measures that we observed in the grey literature (\eg fixing or reruns), we could ask questions why these measures were not adopted.
All the interviews were performed with online calls where we explicitly asked the participants for recording permission. 
The recordings lasted from 26 to 63 minutes with an average duration of 41 minutes.
\subsubsection{Participants}
Our objective was to select practitioners who have experience in dealing with flaky tests in diverse contexts.
This diversity enriches the study and allows us to have a thorough understanding of the practitioner perspective.
To ensure this diversity, we relied on several channels to invite potential participants.

%\paragraph{Contact \& Selection}
We shared our invitation with a brief description of the study objectives on online groups for software engineering practitioners.
For instance, we targeted a group that gathers 265 practitioners that are interested in software testing and continuous integration.
The group members are from large companies like Tesla, Google, Apple, VMWare, Netflix, Facebook, Spotify, etc.
To include participants from other backgrounds, we also targeted groups of practitioners from FinTech companies, average-sized IT companies, and local startups.
%We did not exclude practitioners based on their company size nor their positions.
Following these invitations, we received answers from 19 practitioners who showed an interest in our study.
After exchanges, five participants estimated that their experience is insufficient for the study and did not proceed with the interviews, thus our process ended up with 14 participants.
This number of interviewees is typical in studies that approach similar topics~\cite{tomasdottir2017and,8999994}.
Besides, due to the specificity of the topic, it is very challenging to find other developers that are qualified enough to take part in the study.
We conducted the interviews with the 14 participants and after the analysis, we considered that the collected data is enough to answer our research questions and provide us with theoretical saturation~\cite{glaser2007remodeling}.
Indeed, the three last interviews did not lead to any changes in our analytical template and only provided new formulations for existing categories.

Table~\ref{tab:participants} summarises the profile of our participants (role and years of experience) and their current companies (number of employees, domain, and number of users).
To preserve the anonymity of our participants, we refer to them with code names, we omit their company names, and upon specific request, we also omit the experience and domain.
Our participants have solid experience in software engineering, their experience ranges from 6 to 35 years, with an average of 16 years.
%This strong experience allows us to 
The participants also work in companies that vary significantly in terms of size and domain of activity.
On top of the industrial experience, three of our participants contributed regularly to Open Source Software (OSS) as part of their job or as a side activity.
%We believe that this diversity ...

\begin{table}
\vspace{-0.5em}
\begin{centering}
\caption{A summary of participants' profiles.}
\vspace{-1.0em}
\label{tab:participants}
\resizebox{\columnwidth}{!}{
\begin{tabular}{ccccccc} 
 \cline{2-7}
  & \textbf{\textit{Role}} & \textbf{\textit{
Years}} & \textbf{\textit{Size}} & \textbf{\textit{Domain}} & \textbf{\textit{Users}} & \textbf{\textit{OSS}}  \\  \hline
 P1 & Engineering Manager  & 24 & +1K  & Music & +200M & No \\ \hline
 P2 & CTO & 10 & +10 & Mobility  & - & No \\ \hline
 P3 & Tech Lead & 7 & +200 & Cloud & +30K & Yes \\ \hline
 P4 & QA Consultant & 12 & +2K & FinTech & +190K & No \\ \hline
 P5 & CTO & 14 & +10 & Infrastructure & - & No \\ \hline
 P6 & Staff Engineer & 20  & +1K & DevOPs & - & Yes \\ \hline
 P7 & Vice President & 17 & +200 & Cloud & +30K & Yes \\ \hline
 P8 & Architect & 7 & +5k & Online sales & +70M & No \\ \hline
 P9 & Senior Researcher & 35 & +20 & R\&D & - & No\\ \hline
 P10 & Architect & 30 & +24K & Virtualisation & +500K & No \\ \hline 
 P11 & Senior Engineer & 6 & +10k & - & +500M & No \\ \hline
 P12 & Principal Architect & 23 & +10k & Payment & +200M & No \\ \hline
 P13 & Front-end Developer & 7 & +40 & Banking & - & No \\ \hline
 P14 & Senior Engineer & - & +10k & - & +500M & No \\ \hline
\end{tabular}
}
\end{centering}
\end{table}
\subsection{Analysis}
As our study builds on semi-structured interviews, we relied on the strategy proposed by Schmidt \etal~\cite{schmidt2004analysis}.
This strategy helps with inquiries where a prior understanding of the problem is postulated but the analysis remains open for exploring new topics and formulations.
In the following, we explain the four steps of this analysis.
\paragraph{Transcription}
To prepare the interview analysis, we transcribed the recorded interviews into texts following a denaturalism approach.
This approach allows us to dismiss non-informational content and ensures a full and faithful transcription~\cite{oliver2005constraints}.
For the cases where the interviews were not conducted in English, we transcribed them in the original language and we only proceeded to their translation at the reporting step.

\paragraph{Definition of analytical categories}
The goal of this step is to define the analytical categories that guide our analysis.
In our case the initial categories of interest were:
\begin{enumerate}[wide=10pt,noitemsep,topsep=0pt]
    \item The sources of flaky tests;
    \item The measures for mitigating flaky tests;
    \item The difficulties of dealing with flaky tests; 
\end{enumerate}

After conducting four interviews, we observed that an additional topic that is commonly mentioned by developers is: 
\begin{enumerate}[wide=10pt,noitemsep,topsep=0pt]\setcounter{enumi}{3}
    \item The impact of flakiness;
\end{enumerate}

Based on our preliminary discussions, this topic provided new insights on the effects of flaky tests, as seen by practitioners.
This topic also seemed essential for understanding the efforts dedicated to the mitigation of flaky tests.
Hence, we added this topic to our categories of interest and our interview template.
After setting the analytical categories, we read each participant answer to identify the categories that can be associated with it.
In this process, we do not only focus on the participants' direct answers, but we also consider their use of terms and the aspects that they omit.
For instance, in our analysis of the second analytical category, we consider the measures taken by practitioners but also those that they were not aware of or the ones that they discarded.
On top of that, we carefully analyse developers' answers to context and flakiness questions to spot elements that can help in interpreting their answers.

\paragraph{Assembly of a coding guide}
The objective of this step is to build a guide that can be used to code the interviews.
We assembled the four analytical categories and identified different sub-categories for them based on an initial reading of the interviews.
The sub-categories represent different versions formulated by developers in one  analytical category.
For instance, for the first analytical category, \ie sources of flaky tests, the initial sub-categories were \textsf{Test}, \textsf{Code Under Test}, and \textsf{Environment}.
These sub-categories are not final and can be refined, omitted, or merged along the following step.
For example, the sub-category \textsf{Environment} is later refined to two categories \textsf{Infrastructure} and \textsf{Uncontrollable environment}.

\paragraph{Coding}
We read the interviews to identify passages that can be related to the categories and sub-categories of our coding guide.
This process can be repetitive as every time a new sub-category is identified or refined, we need to read previous texts to ensure that all passages related to it are identified.
To ensure the soundness of this process, two authors coded the interviews separately before comparing their results.
In case of disagreement, the authors discussed their views and opted for a negotiated solution.
Besides this consensual coding, all the authors discussed the coding guide iteratively, to ensure the clarity and precision of the identified sub-categories.

%\subsubsection{}

%% file: study_results.tex
\section{Study Results}
\label{sec:results}
% Table X presents the results of the coding process.
% In this section, we explain these results by detailing each identified sub-category and discussing its relevance with regard to literature.

\subsection{\textsc{RQ1}: Where can we locate flakiness?}
\subsubsection{\textbf{Test}}
8 participants mentioned that the test itself, when poorly written, is a cause of flakiness (P1, P2, P3, P4, P6, P8, P13, P14).
In particular, the participants explained that some tests are by nature difficult to write and prone to flakiness.
For instance, GUI tests were considered as a special cause of flakiness by many participants.
\textit{``The synchronisation points in GUI tests are a major cause of flakiness... We wait for some elements of the web page (\eg button) to proceed to the testing but some other elements could be necessary and lead the test to fail"} (P4).
According to participants, other cases where it is difficult to write flakiness-free tests included time manipulation, threads, statistics, and performance tests.
%Many developers expressed their struggle with writing tests that handle time correctly,
%\textit{``this temporal dimension is not intuitive for developers because they tend to think about the machine as a deterministic black box, while in reality, it is not"} (P7).
%Threading is also perceived as a common source of flaky tests as it requires careful attention from developers to avoid race conditions and timeouts.
%\textit{``Many developers have difficulties while writing multi-threading tests and just tend to use \texttt{sleep()} to handle it. I admit, it is complicated and it is a habit to develop"} (P6).
%For statistics and performance tests, the difficulty resides in the choice of a threshold value since the tested values vary constantly.
%In both cases, the assertion value is difficult to define because the tested values vary constantly, 
%\textit{``assertions in performance tests are catastrophic, the numbers change by $20\%$ because of different noises"} (P6).
%As for statistics, the variation stems from sampling or other randomisation operations, which are difficult to harness and understand.
%\textit{``One time out of ten, we do not have the right sample values and the tests need to tolerate this but without being very permissive and for this we have no magic solutions"} (P6).
Participants also mentioned practices that they observe commonly in tests that lead to flakiness.
P8 described examples of tests that encode variables and properties that are not really useful for the test case and lead to non-deterministic behaviour.
These variables could be related to the system, environment, or time and they can be avoided inside the test code.
% P2 describes a similar issue with tests that handle time-zones while assuming that these variables are constant, which is not always true.
% She suggests that these aspects should be mocked in the tests instead of encoding the real variables.
%Finally, P1 described (cut and test not synchronised)

\subsubsection{\textbf{Code Under Test (CUT)}}
In this sub-category, we consider flakiness that stems from the part of the system that is directly under test.
Surprisingly, only 3 participants mentioned that their flakiness stems from the CUT (P1, P3, P7).
%Some participants even considered it as a more probable source of flakiness than the tests itself,
%\textit{``In my experience, it is pretty rare that the test itself is badly written, most of the time the product is flaky and the tests are just going to find that"} (P1).
The root causes of CUT flakiness are similar to the causes of test flakiness, as examples, the participants mentioned concurrency and time handling.
Interestingly, flakiness in the CUT can have direct impacts on the product reliability and thus developers tend to take it more seriously.
\textit{``If the product itself is flaky, which is happening quite often, then you have got a problem because you actually publish code which is flaky, it breaks one out of three times"} (P1).

\subsubsection{\textbf{System Under Test (SUT)}}
This source of flakiness was mentioned by 9 participants (P2, P3, P4, P5, P6, P7, P8, P12, P14).
Differently from the CUT, this sub-category considers the system as a whole and not only the part under test.
The SUT emerges as a source of flakiness in complex systems where integration tests flake due to failing orchestration between the system components.
\textit{``It only takes one timeout in the communication between two services or other middleware like databases to make a test fail randomly"} (P2).
The failing interactions can be a result of a misunderstanding of the system architecture and its impact on tests,
\textit{``the principle behind micro-services is that every service can fail, so we need to keep that in mind when writing integration tests"} (P2).
The organisational structure can also add to the difficulty of writing stable integration tests as components can be maintained by distinct teams that do not communicate properly.
\textit{``Every team has the impression of working in a sandbox, they would rebase the production or generate a new sequential number and the tests of other teams will flake because of that"} (P8).
Ideally, these dependencies should be documented or formalised and integration tests should account for them.
Yet, P8 confirms that  despite the recurrence of such incidents, developers remain reluctant to invest in their documentation.

%Another facet of flakiness at the SUT level is the concurrency between the system components.
\subsubsection{\textbf{Infrastructure}}
The testing infrastructure is the set of processes that support the testing activity and ensure its stability.
8 participants considered that their tests were flaky because of an unstable or improper testing infrastructure (P1, P4, P5, P6, P10, P11, P12, P14).
For instance, P5 explained that most of their flaky tests were caused by a lack of resources, \textit{``the test is getting throttled because we do not have enough CPU or memory quota for our database"}.
%Similarly, P6 highlighted the effect of the poor setup of their continuous integration, \textit{``our CI is very slow because of an overload or a noisy neighbour that will make the tests timeout"}.
P12 showed how flaky tests can emerge from a mismatch between the product design and its usage in the testing infrastructure, \textit{``a single data source that would, in production, be used by only one user, now is used by several tests that may override each other's data"}.
%Another example of fragile infrastructure is presented by P4, who described how their UI tests flake frequently because of the unstable staging environment \textit{``we have brutal deployments in the staging and we frequently find ourselves incapable of testing correctly because of it"}.
When flaky tests are caused by poor infrastructure, participants express more struggle in detecting and fixing them as the search space is broader and programmers are not always qualified for these tasks, \textit{``CI issues are not like race condition where we can have a clear solution for it, this is difficult because it can be different things''} (P6).

\subsubsection{\textbf{Environment}}
11 participants explained that tests can flake because of external factors (P1, P2, P5, P6, P7, P8, P9, P10, P11, P13, P14). 
This source of flakiness differs from the infrastructure by considering all factors that developers cannot or should not control.
One common example of the environment is the hardware on which developers have almost no control, \textit{``sometimes one batch of RAM sticks has an unidentified problem and the test is failing because of it"} (P7). 
The underlying Operating System (OS) is also subject to various changes that make it unpredictable and therefore a potential source of flakiness.
One example of such cases is given by P1: \textit{``if we test the app on devices, then we rely on some iPhone being up and if it decides to upgrade its OS at the exact same time then we have a problem"}.
On top of the OS, tests can always be impacted by cumulative states of the machine that developers do not account for, \eg firmware versions, memory state, and access to the internet.

The impact of the environment is particularly perceptible on GUI tests since they run on different web browsers that are prone to frequent changes.
%\textit{``Even if you put efforts into making a GUI test deterministic, at the moment when you run it on different frameworks (Chrome, Firefox, Internet Explorer, etc) you lose all control and that is why GUI testing is hard"} (P5).
Similarly, developers may need to write acceptance and integration tests that depend on external resources that are hardly controllable.
\textit{``I work on a command-line interface that wraps packages from different providers, it seems simple but there are always random changes"} (P7).

It is worth noting that the distinction between infrastructure and environment may depend on the software, test type, and the choices of the practitioner.
Some developers can consider aspects like the OS state as part of their infrastructure and control it to ensure the reliability of their tests, whereas others choose to ignore it.
Likewise, aspects that seem external and futile for unit or integration tests, \eg firmware, must be considered and controlled as part of the infrastructure of performance tests.
%\textit{``Colleagues working on performance tests had to tune details like the processor and BIOS to avoid noise and ensure that their tests make sense"} (P6).

\subsubsection{\textbf{Testing framework}}
Two developers found that the testing framework can lead to flakiness (P1, P7).
This issue can arise when the framework is written or customised by the developers themselves, which makes it less stable than other widely used frameworks.
Another possible issue is the mismatch between the testing framework and the CUT.
%For instance, the CUT can be asynchronous while the testing framework is not. 
%This forces developers to use synchronisation points that synchronise the tests but also make them prone to flakiness.
This can occur when the framework is not adapted to the type of tests or to the application domain.
P7 describes a similar case: \textit{``We used a Cassandra cluster (NoSQL) and we tried to test the database consistency rules. This generated many flaky tests. Instead, we should have used a more delicate testing framework to write serialisation tests and produce consistency edge cases"}.

\subsubsection{\textbf{Tester}}
Two participants believed that developers and testers can constitute a source of flakiness (P4, P5).
This is possible for manual tests where the tester actions are part of the test execution.
Indeed, being manual makes tests rely on human behaviour, which is less deterministic and more failure-prone.
Hence manual tests can flake because of variations in the tester actions.
Besides, the tester's misunderstanding of the requirements and the SUT can be another point of failure.
\textit{``The person running the tests does not always have a correct and precise idea of the behaviour expected from the system and this affects the test outcome"} (P4).

% \begin{table}[ht!]
% \begin{centering}
% \vspace{-0.5em}
% \caption{The sources of flakiness.}
% \label{tab:RQ1}
% \vspace{-0.5em}
% %\resizebox{\columnwidth}{!}{
% \begin{tabular}{ccc} 
%  \hline
%   \textbf{\textit{Flakiness Source}} & \textbf{\textit{\#Participants}} & \textbf{\textit{\%Participants}}  \\  \hline
 
%  \textit{Environment} & 11 & 78\%\\
%  \textit{SUT} & 9 & 64\%\\
%  \textit{Infrastructure} & 8 & 57\%\\
%  \textit{Test} & 8 & 57\%\\
%  \textit{CUT} & 3 & 21\%\\
%  \textit{Testing framework }& 2 & 14\%\\
%  \textit{Tester} & 2 & 14\%\\ \hline
% \end{tabular}
% %}
% \end{centering}
% \vspace{-0.5em}
% \end{table}

\paragraph{\textbf{Discussion}}
% Table~\ref{tab:RQ1} summarises the sources identified in our interviews.
% It is worth noting that one participant can mention multiple sources of flakiness. 
According to our participants, flaky tests stem frequently from the external factors of the environment, the interactions of the SUT, and the testing infrastructure.
Flakiness is not limited to the test and CUT and the studies on this topic should consider and leverage all these factors when addressing flaky tests.
Our analysis also shows that besides the well-established root causes of flaky tests, \eg concurrency and order-dependency, the size and scope of the test are important flakiness factors.
GUI and system tests are more prone to flakiness, yet, our understanding of flakiness in these types of tests remains limited and we still lack techniques that adapt to these specific tests. 

\subsection{\textsc{RQ2}: How do practitioners perceive the impact of flakiness?}

\subsubsection{\textbf{It wastes developers' time}}
10 participants considered that flaky tests waste developers' time (P2, P4, P5, P6, P7, P8, P9, P11, P12, P14).
When developers observe flaky failures, they have to invest time and effort in investigating the root cause before realising that it is a false alert.
%\textit{``Developers waste so much time checking false positives instead of developing"} (P6).
% The time and efforts invested in analysing flaky failures increase in accordance with the test scope.
% Hence, the analysis of GUI tests, which have a larger testing scope, tend to be more time and effort consuming. 
%\textit{``the cost is tremendous because we do not only mobilise testers, but all developers who could be concerned by the failure"} (P4).
Besides the time wasted on investigating false alerts, our participants affirmed that discussions about flaky tests are also costly.
% In particular, when flaky tests remain unaddressed, developers are frequently required to explain these false alerts to other stakeholders.
\textit{``It was ok when we were a team of five and everyone knew that the test is flaky. But as the startup grew, it became expensive and we found ourselves constantly explaining to other developers that these are not real failures"} (P7).

\subsubsection{\textbf{It disrupts the CI}}
7 participants mentioned that their flaky tests disrupt the continuous integration process (P2, P3, P4, P8, P10, P11, P13).
This impact arises from the pace of modern development life cycles and its extent is proportional to the releasing frequency. 
\textit{``Flakiness would never be an issue if we released once every two weeks. But in a CI today with 400 deliveries per day, disruptions waste so much time"} (P2).
%The same participant qualified these disruptions as the most detrimental effect of flakiness, \textit{``computation and development time are not relevant in our case, developers would always waste time somewhere. The cost that interests us is the time wasted for the release"}.
The disruption can be particularly costly when the release is critical, \eg an urgent hot-fix that could be delayed because of flaky failures.
Disruptions also affect the developer's ability to develop confidently because the CI, which is supposed to guard the code quality, is halted.
For instance, P3 who worked on a project where tests were failing at the beginning of each month (because of improper time quantification) stated: \textit{``five days a month, the Jenkins of this project was red so I couldn't develop on the project and be sure that my work is not breaking anything at the time"}.

\subsubsection{\textbf{It affects testing practices}}
6 participants observed that flakiness affects the testing practices in their teams (P1, P6, P7, P8, P12, P13).
In particular, they explained how developers lost confidence in their capacity to write tests, 
%\textit{``it is really discouraging, and they start to believe that integration testing is very complex and they are unable to test or fix flaky tests"} (P8).
According to P12, in the worst-case scenario, developers are repelled and would write fewer tests to have fewer problems.
In a phenomenon similar to the broken window theory, P1, P7, and P12 described how developers are more inclined to introduce and accept flaky tests in a system that is already flaky.
\textit{``As the suite is unreliable, it opens the door for more flaky tests"} (P7).
Ultimately, the accrual of flaky tests pushes development teams to adapt their testing strategies: \textit{``flakiness tends to accumulate in the system, and at some point, it becomes so large that companies may look for completely different solutions, like using more unit testing"} (P12).
The impact on testing practices is not only related to flakiness but also to the general software quality, \textit{``the more flakiness it is, the greater the acceptance of less than ideal test coverage, and that leads to a degradation of the software quality"} (P12).

\subsubsection{\textbf{It undermines the system reliability}}
5 participants highlighted the impact of flaky tests on the reliability of both tests and the SUT (P1, P3, P6, P7, P8).
The false alerts raised by flaky tests confuse developers and make them question the suite's ability to detect faults accurately.
%\textit{``Flakiness discourages development teams and they lose trust in tests"} (P8).
Consequently, developers can disregard test results, which may lead to the introduction of bugs, \textit{``if you do not fix flaky tests, people will start ignoring them and then they will introduce real bugs in the product"} (P1).
%As a result: ignoring potential bugs
Similarly, the non-deterministic test outcomes cast doubts on the reliability of the system under test.
%P7 explains the impacts of these doubts: \textit{``since we work on a billing system, these incidents stressed the team and developers always had doubts"}.
This doubt is all the more important in open source projects where newcomers can be repelled by inexplicable flaky failures.
P6 who worked on a large open-source project stated: \textit{``new contributors see CI failures, they do not know it is flakiness and it gives them the impression that the project is not well maintained so they do not even rerun the tests, they just give up"}.

\subsubsection{\textbf{It disguises bugs}}
Two developers explained that flakiness can hide buggy features (P1, P6).
In some cases, the non-deterministic behaviour stems from a bug in the product, but as developers believe it is a flaky test, they disregard it without further inspection.
\textit{``People ignore the flaky test results because it is just a flake, except it is an actual problem in a product"} (P1).
%The same developer affirmed that these cases were not rare in their system: \textit{``we had a problem because we actually published code which is flaky to our customers, and people don't usually like that one out of three times the feature breaks"}.
Interestingly, we witnessed first-hand the confusion between buggy features and flaky tests while performing the interview with P9.
The participant was providing an example of non-deterministic test failures that were caused by memory issues, and when asked about how these flaky tests were detected, she replied: \textit{``they appear when they are in the customer premises"}.
After the customer complaint, the participant reran the test that covers the buggy code multiple times and reproduced the bug.
In this case, the test has indeed a non-deterministic outcome, but addressing it as a flaky test (false alert) is inappropriate because the failure is real.
Furthermore, the more flakiness is prevalent in a test suite the more developers are inclined to overlook non-deterministic system failures.
\textit{``The most important is that it actually makes people think they can introduce bugs in the form of flaky bugs in a product and get away with it"} (P1).

\paragraph{\textbf{Discussion}}
Our results confirm the impact of flakiness in terms of development time and CI obstruction.
Moreover, our analysis shows that the accrual of flaky tests affects the testing practices negatively as developers become repelled by testing and more lenient toward testing standards, which eventually leads to a degradation of the system quality.
Our participants also raised the issue of system buggy non-deterministic features that are falsely labelled as test flakiness and therefore disregarded and shipped to end-users. For future studies, this shows the necessity of distinguishing the sources of flakiness and addressing them accordingly.
%add
\subsection{\textsc{RQ3}: How do practitioners address flaky tests?}

\begin{table*}[htbp]
\vspace{-0.6em}
\centering
\caption{The number and percentage of grey literature articles and interviews for each mitigation measure.}\label{table:strategies}
\vspace{-1.1em}
\begin{tabularx}{\textwidth}{lXcccc}
 \cline{2-6}
     \multicolumn{1}{c}{} &  \textbf{\textit{Strategy}}  & \textbf{\textit{\#GL}} & \textbf{\textit{\%GL}} & \textbf{\textit{\#Int.}} & \textbf{\textit{\%Int.}}\\  \cline{2-4}
    \hline
    
    \multirow{4}{*}{{\textbf{Prevent}}} &  \textbf{Setup a reliable infrastructure} with processes properly adapted to the testing activity.
     & 4 & 11\% & 9 & 64\% \\  %\cline{2-4}
     & \textbf{Define guidelines} that should be respected when writing tests and enforced through reviews.   & 5 & 13\% & 9 & 64\%\\  %\cline{2-4}
    & \textbf{Limit external dependencies} by mocking connections, services, and dependencies. & 9 & 24\% & 1 & 7\% \\  %\cline{2-4}
    
    & \textbf{Customise the testing framework} to avoid flaky features. & 4 & 11\% & 1 & 7\% \\  \hline

    \multirow{5}{*}{{\textbf{Detect}}} 
    
    & \textbf{Rerun} the failing test multiple times to check if it is a real failure or a flaky one.  & 14 & 37\% & 7 & 50\% \\  %\cline{2-4}
    & \textbf{Manually analyse} the failure message and trace to determine if the test is flaky.  & 17 & 45\% & 3 & 21\% \\ %\cline{2-4}
    & \textbf{Check the test execution history} to distinguish flaky failures from real failures.  & 8 & 21\% & 2 & 14\% \\ %\cline{2-4}
    & \textbf{Proactively expose} test flakiness before it manifests in the CI.   & 5 & 13\% & 2 & 14\%\\ 
    %\cline{2-4}
    & \textbf{Compare test coverage} to the modifications of the commit under test to identify flaky failures.  & 2 & 5\% & 1 & 7\% \\ 
      \hline

    \multirow{5}{*}{{\textbf{Treat}}} & \textbf{Fix} the root cause of flakiness to remove the non-deterministic behaviour. & 15 & 39\% & 7 & 50\% \\  %\cline{2-4}
    & \textbf{Ignore} flaky tests that are not common or costly (based on the flake rate and periodicity). & 2 & 5\% & 5 & 36\%\\  %\cline{2-4}
    & \textbf{Quarantine} flaky tests by isolating them from the blocking path that commands the CI. & 7 & 18\% & 4 & 23\% \\  %\cline{2-4}
    & \textbf{Remove} the test permanently from the suite. & 2 & 5\% & 4 & 23\% \\  %\cline{2-4}
    & \textbf{Document} flaky tests in databases, issues, alerts, or internal reports. & 8 & 21\% & 3 & 21\% \\  \hline
    
     \multirow{2}{*}{{\textbf{Support}}} & \textbf{Monitor and log} system interactions and test outcomes. & 8 & 21\% & 9 & 64\% \\  %\cline{2-4}
     & \textbf{Establish testing workflows} that protect the CI. & 2 & 5\% & 4 & 23\% \\ \hline
\end{tabularx}
\vspace{-0.7em}
\end{table*}
Table~\ref{table:strategies} summarises the measures identified in our GLR.
The columns \textit{\#GL} and \textit{\#Int.} report the number of times where the measure was mentioned in grey literature and interviews, respectively, while the columns \textit{\%GL} and \textit{\%Int.} report the percentages.
The full results summary is available within our artefacts~\cite{artefacts}.
%In the following we explain these measures and we report on the explanations gathered from the interviews.

\subsubsection{Prevention measures}
This represents all proactive practices that aim to prevent the introduction of test flakiness.

\paragraph{\textbf{Set up a reliable infrastructure}}
Grey literature articles that embraced prevention measures estimated that a proper setup of the testing infrastructure is necessary for avoiding flaky tests.
%Naturally, this setup depends on the project and test type, but, many actions are commonly taken by practitioners to ensure the reliability of their infrastructure.
Several practitioners adopted hermetic servers, \aka mock servers, where tests can be run locally without the need to call external servers~\cite{TestStab71:online,Flakytes0:online,Howtofix72:online}.
Some articles also stressed the importance of using containers to ensure that the testing environment is clean when the tests are run~\cite{Effectiv86:online}.
9 interviewees reported the adoption of similar practices to ensure the stability of their infrastructure (P1, P3, P4, P5, P6, P10, P11, P13, P14).
P6 explained that they rarely observe test failures caused by infrastructure or environment thanks to their use of virtual machines.
\textit{``The virtual machine is started for the tests and destroyed just after ... all our tests are reproducible"} (P6).
%In the case of P5, the use of containers was motivated by observations of recurrent flaky systems tests, \textit{``I witnessed that tests were failing due to cumulative changes on the system so now the containers restart every time and the environment is cleaned"}.
P4 mentioned \textit{pre-tests}, a form of sanity checks, as another solution to infrastructure flakiness.
%These \textit{pre-tests} ensure that all the resources needed for the test suite are available before running it.
\textit{``If we have 5 APIs involved, the pre-tests check that these APIs are up, otherwise the test is not run"} (P4).
%With respect to the motivation behind this measure, the 9 participants explained that this measure seemed necessary for their integration testing or emerged following costly flakiness issues.
%\textit{``Companies only make such investments when overwhelmed by flaky tests"}, stated P4.

\paragraph{\textbf{Define testing guidelines}}
One guideline that was recurrently mentioned is following the testing pyramid principles~\cite{FlakyTes87:online,Managing72:online,Effectiv86:online}.
These basic principles force developers to respect the scope of each test type and avoid flakiness.
%The pyramid also specifies the proportions of each test type and privileges smaller tests, in terms of scope, over larger tests like GUI tests.
The proportions of each test type shall also be respected to avoid the \textit{Ice cream cone} and the \textit{Cupcake} anti-patterns~\cite{Introduc67:online}, where the number of GUI tests, which are a main source of flakiness, is exaggerated.
%Indeed, a common guideline is to privilege smaller tests in terms of scope and size.
%At Google, practitioners have even defined new size metrics to measure tests in terms of used resources and they show that smaller tests are less prone to flakiness~\cite{?}.
Interestingly, only two of our interviewees (P11 \& P14) confirmed that their teams defined explicit guidelines to prevent flakiness.
P14 considered that thanks to these explicit guidelines, she rarely encounters flaky tests in her product, \textit{``with the investment that was done in the guidelines and tooling, now we are able to cope with flakiness"}.
The other participants suggested that the absence of explicit guidelines in their companies is due to the lack of maturity (P8).
%Furthermore, P1 argues that it is unnecessary to define such guidelines and prefers a bottom-up approach, 
%\textit{``we let people learn from their own environment and they realised by themselves to avoid mixing synchronous and asynchronous code"}.
However, many participants affirmed that with experience their teams had developed testing practices to avoid flaky tests (P2, P3, P4, P6, P7, P10, P13).
These practices are similar to the ones identified from the grey literature.
They focus on the test scope and size and they address common flakiness sources like concurrency and time manipulations.
In order to enforce these good practices, the participants relied on code reviews.

\paragraph{\textbf{Limit external dependencies}}
This practice is more relevant for unit tests, which are supposed to test narrow parts of the systems, than integration or GUI tests, which have to interact with other components.
The analysed articles explain that some practitioners keep useless dependencies in their unit tests, which lead to flakiness~\cite{Managing72:online,TestStab71:online}.
P3 mentioned that in order to avoid environment flakiness, her team tries to mock external services, use test doubles, and prefer in-memory resources (\eg database and file system).

\paragraph{\textbf{Customise the testing framework}}
Sudarshan \etal~\cite{Nomorefl8:online} explained how they built their own testing framework so they can test critical aspects like time and concurrency without introducing flakiness. 
In some cases, practitioners customise the testing framework to disable features like animations in web and mobile applications, which are commonly connected to flaky tests~\cite{TestStab71:online}.

%add manual checks need for static analysis

\subsubsection{Detection measures}
This category groups all actions taken by developers to identify flaky tests.

\paragraph{\textbf{Rerun}}
Based on our GLR, reruns are the most common and intuitive way of identifying flaky tests despite their computation cost.
Even other measures and mitigation steps, \eg debugging and reproduction, require multiple test reruns.
To maximise their chances to observe flakiness and minimise the number of reruns, the reruns can be performed in different environments (local machine, CI, etc) and with different settings~(P4).
Some participants advocated the effectiveness of reruns especially for infrastructure and environment flakiness (P1, P2, P4, P5, P10, P11, P12).
%For instance, P5 considered that when her flakiness issues stemmed from a polluted system, the \textit{``restart and rerun"} was always effective.
Nevertheless, P1 warned about the consequences of solely depending on reruns to deal with flakiness, \textit{``with reruns, you do not understand the issue and you can ignore actual problems"}.

\paragraph{\textbf{Manually analyse test outcome}}
When even reruns are not possible or useful, developers manually analyse the execution trace to determine if the test is flaky or not~\cite{Preventi83:online}.
In the case of GUI tests, practitioners rely particularly on the screenshots recorded during the test run~\cite{Flakytes0:online,FlakyTes55:online,Selenium34:online}.
P2, P4, and P8 affirmed that they prefer going through manual analysis before trying reruns or other detection techniques.
In the case of P8, this choice is due to system specifications that make rerunning the same test in the exact same conditions impossible. 
%\textit{``For example, if we already have pictures of the product, it is technically impossible to take them again in our system, so the test cannot be rerun"} (P8).

\paragraph{\textbf{Check test history}}
Some practitioners keep a record of the test execution history, \ie all test passes and fails for each build.
When a suspicious test failure is observed, developers inspect these records to check if the test has already shown a random behaviour.
Palmer~\etal~\cite{TestFlak61:online} argue that when these records are visualised they can help developers in distinguishing flaky failures easily and thus gain a lot of investigation time.
P11 and P14 described a system in their company, which relies on the execution records to score tests.
Based on the past passes and failures, a test receives a flakiness score that expresses the probability for this test to be flaky.
P14 described how these scores helped her when a flaky test manifested, \textit{``it is very good when it tells that it is 90\% flaky and you can just go on with your day knowing that it's because of flakiness"}.

\paragraph{\textbf{Expose}}
As explained in RQ2, when a flaky failure occurs in the CI, it disrupts the work progress and wastes developers' time and efforts.
For these reasons, some practitioners attempt to reveal flaky tests before CI failures~\cite{Amachine72:online,TestFlak61:online}.
In this case, new tests are rerun several times to ensure that they are stable, before adding them to the main test suite.
Among our interviewees, only P1 and P4 reported adopting this practice in their companies.
\textit{``Before committing the test, you should run it a thousand times (counting different configurations and device types) and it must be a thousand greens (passes)"} (P1).
%Other participants judged these proactive measures as unnecessary for their situations. 
%P1 considered that some of their order-dependent tests could have been detected if they proactively rerun the test while shuffling the execution order.
%However, she considered such outcome is not worth the needed investment, \textit{``given that we do not have a lot of flakiness appearing, we do not need to to push a lot"}.

\paragraph{\textbf{Leverage test coverage}}
When practitioners suspect that a test failure is flaky, they compare the coverage of the failing test to the modifications performed by the commit that triggered the build.
If the intersection between these two is empty, the test is considered flaky.
This process can be performed manually by developers (P14) or automatically using tools like DeFlaker~\cite{bell_deflaker_2018}.
However, P14 explains that, due to hidden dependencies between projects, this technique is not always effective.

\subsubsection{Treatment measures}
This presents actions taken by practitioners to deal with flay tests that manifested.

\paragraph{\textbf{Fix}}
In theory, every identified flaky test should be fixed at some point.
However, according to practitioners, this point is rarely reached because the fix depends on two challenging steps, reproducing the flaky failure and determining its root cause (\textit{cf. } RQ4).
For this reason, many flaky tests remain unaddressed or removed.
Interestingly, some participants affirmed that fixing flaky tests is easy when the root cause is known (P2, P10).
P3 also affirmed that once the flaky test is understood, it was only a matter of resources to fix it.
% P1, P3, P6, P11, and P13 insisted that, similarly to bugs, flaky tests should be fixed as soon as possible.
% \textit{``I prefer debugging and fixing the flaky test immediately instead of having to debug a race condition in my production, which is a nightmare"} (P6).

\paragraph{\textbf{Ignore}}
Naturally, ignoring flaky tests is not commonly recommended in the grey literature (only 2 articles).
Yet, 5 interviewees recalled situations where flaky tests were intentionally left unaddressed (P2, P3, P6, P7, P10).
For P3 and P7, this was in a case where all team members were aware of the test flakiness and considered that the test is useful, so they did not isolate or remove it, but did not have enough time or resources to fix it.
%In this situation, as the flaky test is also left undocumented, it entailed communication costs and wasted developers' time as shown in RQ2.
For P6 and P10, this choice is motivated by the severity of the flaky test, \ie the flake rate.
\textit{``If the test has a very low flake rate, it is not really worth the investigation"} (P10).

\paragraph{\textbf{Quarantine}}
According to our GLR, quarantining flaky tests is one of the most common measures among practitioners.
While in most cases, the isolation in quarantine is performed manually by developers when they identify a test as flaky, in some cases this process is more sophisticated.
An article from Fuchsia explained how they designed an automated workflow where flaky tests are automatically identified and removed from the commit queue~\cite{Flakytes54:online}.
This workflow comprises a benchmark that evaluates the fixed flaky tests before reinserting them in the integration suite.
By lack of better solutions, this evaluation relies on reruns.
The adoption of the quarantine is less popular among our interviewees (P1, P4, P7, P10).
Indeed, even participants who affirmed that they isolated their flaky tests, raised several questions about the side effects of this practice.
P1 suggested that developers can abuse this practice, \textit{``it's a dangerous way to go because then suddenly the number of tests goes down"}.
P6 went further and considered that the quarantine is a bad practice because it implies that a potential bug is being disregarded without further investigation.
\textit{``You move the problem from the developer, who will not see the flaky failures anymore, and you transfer it to the user who may deal with a bug"} (P6).
% In the same vein, P3 considered that even a flaky test has an important value (fault detection) and should only discard it if it is redundant or its flake rate is so high.
\paragraph{\textbf{Remove}}
When a flaky test is hard to reproduce, debug, or fix, many practitioners recommend to remove it completely from the system to avoid its negative effects~\cite{Flakytes54:online,FlakyTes82:online,ThinkLik1:online}.
P1, P2, P7, and P14 affirmed that if a flaky persists and they are unable to address they choose to remove it.
\textit{``I would rather remove the flaky test from the codebase because of its cost"} (P2).

\paragraph{\textbf{Document}}
The documentation of flaky tests is performed for different purposes.
The most basic being informing other developers that the test is flaky so they know how to react to its failures.
The documentation is also helpful for the reproduction and debugging of flaky tests as it keeps logs, memory dumps, system states, screenshots in GUI tests, etc~\cite{flakytes70:online}.
Finally, keeping track of all flaky tests is helpful when building a system that relies on execution history to detect flaky failures.
Indeed, three interviewees affirmed that their internal systems relied on flaky tests that were documented in the past (P10, P11, P14) to guide developers when a test fails.

\subsubsection{Support measures}
This includes actions that are likely unrelated to test flakiness but are critical for addressing flaky tests.

\paragraph{\textbf{Monitor and log}}
9 interviewees explained that when addressing a flaky test they rely mainly on the data logged by their monitoring system (P1, P6, P8-P14).
P1 explained their advanced log analysis, which automatically suggests the root cause of the failure, \textit{``we have a probe that can identify those root causes of flakiness"}.
Regarding, the effect of this monitoring and analysis on their productivity, P1 added: \textit{``it takes years to do it right, but it is extremely powerful"}.
P11 and P14 explained that the test logs assist their flakiness prediction system.
Furthermore, P6 and P10 showcased the importance of monitoring by affirming that their decisions are always guided by the flake rate, a test score that is calculated by monitoring and analysing test outcomes for periods of time.

\paragraph{\textbf{Establish testing workflows}}
For complex software systems, practitioners can design advanced testing paths that organise tests based on their criticality for the integration~\cite{Selenium34:online,WeHaveAF52:online}.
In these scenarios, due to computation costs, the blocking path, \ie the set of tests that decide in the CI, does not include all tests.
4 interviewees suggested that these workflows can be leveraged to protect the blocking path from flaky tests (P1, P10, P11, P14).

\paragraph{\textbf{Discussion}}
Our analysis shows that on top of the typical detection and treatment measures, developers take actions to prevent the introduction and manifestation of flaky tests.
Interestingly, this prevention relies mainly on the setup of the infrastructure and the establishment of guidelines.
To the best of our knowledge, these two tasks were not identified by prior studies and none of the literature techniques supports them.
Similarly, our results emphasise the role of supporting measures like logging and monitoring in the accomplishment of critical mitigation steps like detection and fixing.
The study of Lam~\etal~\cite{Lam2019RootCausing} has already shown that logs can be used to automatically spot the root cause of flakiness. 
Other studies should follow the same path and benefit from monitoring and log analysis to improve flakiness detection and prediction.
%Detection is still based on manual analysis:> unfortunate!

%Observe
%Compare to related work
%Conclude

% \begin{tcolorbox}[boxsep=1pt,left=2pt,right=2pt,top=2pt,bottom=2pt]
% \textbf{\textsc{RQ3:}}
% \end{tcolorbox}

\subsection{\textsc{RQ4:} How could mitigation measures be improved with automation tools?}

\subsubsection{\textbf{Root cause identification and reproduction}}
8 participants expressed their struggle while reproducing and debugging flaky tests (P1, P2, P3, P4, P7, P9, P10, P11).
These two tasks are tightly coupled because reproducing a flaky failure generally requires a minimal understanding of the root cause.
P4 explained that the difficulty of these tasks is due to the multitude and variety of potential factors of flaky tests, both in terms of root causes and sources (from the test itself to complete external factors).
P11 added that the broadness of factors is particularly relevant for SUT flakiness: \textit{``Trying to figure out among 8 to 10 services what is the actual culprit of flakiness is the challenging part"}.
For all the participants, except P1, the reproduction and debug are currently performed manually, which is time and effort consuming.
P7 affirmed that simple reruns are not always effective for reproducing and more advanced solutions are necessary, \textit{``we need tracking tools to help us reproduce flaky tests"}.
In the same vein, P4 said that even when logs are available, a lot of assistance is still required to help developers isolate the root cause and reproduce flaky tests.

\subsubsection{\textbf{Monitoring and log analysis}}
7 participants suggested that managing flaky tests would be easier if they were equipped with tools to monitor the testing activity and analyse the generated logs (P3, P4, P6, P8, P9, P12, P13).
These two tasks are coupled because an automated analysis is critical to benefit from the data collected by the monitoring process.
Indeed, P4 said that their GUI testing system produces overwhelming amounts of logs and yet it is impossible to manually draw insightful information from them.
The analysis of such data can help developers to: 
\begin{itemize}[wide=10pt,noitemsep,topsep=0pt]
    \item\textbf{Predict flaky tests}: As shown in \textsc{RQ3}, analysing the logs of test history is useful for predicting flakiness and assisting developers when a flaky failure occurs.
    \item \textbf{Identify the source or root cause}: \textit{``For debugging GUI tests, traces of all the called APIs can help in isolating the root of failure"} (P4).
    \item \textbf{Evaluate the flake rate:} In \textsc{RQ3}, we showed that the flake rate monitoring gives a fine grained assessment of flaky tests and therefore guides the mitigation strategies, \eg ignore flaky tests that flake rarely.
    \textit{``This monitoring would help us to debug and find the changes that led to increasing the flake rate"}, stated P6 who explained that these tasks are currently performed manually.
\end{itemize}

\subsubsection{\textbf{Test validation}}
\textsc{RQ3} showed that following testing guidelines is a key measure for preventing test flakiness. 
Yet, according to 9 participants, the process of enforcing these guidelines still relies on manual reviews, and it could be assisted with:
\begin{itemize}[wide=10pt,noitemsep,topsep=0pt]
\item \textbf{Static analysis}: P10 described how preventing flakiness through code reviews can be redundant, \textit{``I keep rejecting tests that have sleep() statements"}, and suggested that a simple static analyser could help in this regard.
P4 described a similar situation with GUI testing reviews and affirmed that \textit{``advanced static analysis could help to identify potential problems"}.

\item \textbf{Variability-aware reruns}: 
%P1 and some reviewed articles~\cite{TestFlak61:online} already showed that reruns could help in assessing the test before integrating it in the blocking path.
P4 mentioned that she currently tests the scripts of GUI tests manually: \textit{``I test the script by crashing the browser and observing the outcome. This avoids pushing flaky tests that block the quality gate"}.
P6 emphasised the need for tools that automate such procedures: \textit{``it would be great to have a tool that stress tests the tests to ensure their stability"}.
Indeed, the manual test validation could be assisted with variability-aware reruns that account for different configurations, inputs, and system states (\eg \cite{WongMLK18}).
These variations can build on the known causes of non-determinism (\eg random inputs and the system resources) to expose, detect, and reproduce flaky tests. 

\end{itemize}

\paragraph{\textbf{Discussion:}}
Our results confirm previous observations~\cite{eck_understanding_2019,Lam2019RootCausing,Lam2020UnderstandingReproducibility} and show that reproducing and debugging flaky tests remain the most challenging tasks for developers.
%Cite other works on the localisation of the root cause
Furthermore, our analysis accentuates the need for techniques and tools that monitor and analyse the system states to assist the prediction, debugging, and evaluation of flaky tests.
This need is particularly relevant if we consider the results of \textsc{RQ1}, which suggested that flakiness can stem from the system interactions and factors that are external to the source code.
Indeed, trace analysis could be a powerful tool that complements the current detection and prediction approaches, which rely mainly on the source code~\cite{bell_deflaker_2018,lam_idflakies_2019,Pinto2020,King2018,Bertolino2020}.
Our results also show that a more fine-grained analysis of flaky tests, using the flake rate, can be more insightful for developers. 
This aligns with the works that suggested that every test is potentially flaky~\cite{harman2018start}, and research studies should focus on (or at least consider) the level of flakiness instead of classifying tests as flaky and non-flaky.
Finally, our participants expressed the need for automating the quality assessment of software tests through static analysis and variability-aware reruns.
In particular, techniques that rerun tests with different configurations or inputs, like Shaker~\cite{Silva2020} and FLASH~\cite{dutta_detecting_2020}, seem very promising if we consider the role of external factors on flakiness.

%% file: threats.tex
\section{Threats to Validity}
\label{sec:threats}
%\paragraph{Transferability} 
A possible threat to the generalisability of our study is the number of participants. Unfortunately, %It would have been preferable to have a larger set of participants, but
due to the specificity of the topic, it was challenging to find developers qualified to take part in the study. We tried to ensure the quality of our results by only considering practitioners with relevant experience (with flakiness in particular and testing in general). The experience of our participants ranges from 6 to 35 years, with an average of 16 years.
Our participants also constitute a diverse set of roles, company sizes, and application domains.
Moreover, %after analysing the interview transcripts, we estimated that 
the collected data are enough to answer our research questions and provide us a theoretical saturation~\cite{glaser2007remodeling}.

%\paragraph{Credibility} 
A potential threat to the credibility of our findings could be the credibility of the analysed materials as we relied on grey literature and interview transcripts.
In grey literature, we followed the quality assessment guidelines of Garousi \etal~\cite{garousi2019guidelines}, which were specifically designed for such purposes.
In interviews, we communicated the study objectives to the participants and clearly explained that the process is not judgemental.
Moreover, we formulated our questions to target the practitioner experiences and observations.% instead of the knowledge that could be acquired from external sources.

%\paragraph{Confirmability}
A potential threat to the confirmability of our results is the accuracy of the analysis of the transcripts.
To mitigate this threat, two authors coded the interviews separately before comparing their results.
Afterwards, the authors discussed their disagreements and opted for negotiated solutions.
Besides the consensual coding, all the authors discussed the coding guide iteratively, to ensure the clarity and precision of the identified sub-categories.

%% file: conclusion.tex
\section{Implications}
\label{sec:conclusion}
We presented an analysis of flaky tests and their mitigation strategies through a qualitative analysis of 14 practitioner interviews and a grey literature review.
%We performed a qualitative analysis of 14 practitioner interviews and a grey literature review to identify the sources, impacts, mitigation measures, and challenges of test flakiness in practice.
Our study shows that the analysis of flaky tests must consider the whole testing ecosystem and it should not be limited to the test and code under test.
We also highlight a broader impact of flakiness on the testing practices and the overall system quality than what had been presented by previous work. 
Finally, we synthesise 16 measures adopted by practitioners to mitigate flakiness and we identify automation opportunities within them. 
These results open an avenue for future work:

\begin{itemize}[wide=10pt,noitemsep,topsep=0pt]
    \item Flakiness stems mainly from the interactions between system components, the testing infrastructure, and uncontrollable external factors. 
    Future studies can leverage monitoring and log analysis to propose techniques that assist practitioners in addressing flakiness.
    
    \item The establishment of simple testing guidelines, \eg recommendations on test size, external resources, and assertion thresholds, is a key measure for preventing flaky tests.
    Future studies can decrease the manual effort expended in enforcing such guidelines by providing static analysis tools and code review processes.
    
    \item Future work can leverage variability-aware reruns \cite{WongMLK18} and fuzzy testing to effectively expose and reproduce flaky tests.
    Such techniques can help in automating the current manual test validations performed by practitioners.
    
    \item Given the frequency of flaky tests and the cost of their mitigation, practitioners rely on the flake rate to adapt their strategies.
    Future work should account for this indicator when assessing flaky tests and leverage it in their automated solutions.

    \item Some practitioners may falsely label buggy and non-deterministic features as flaky tests, and thus ignore them and treat them as false alerts.
    Future studies should further investigate the impacts of such confusions.
    
    \item Due to the difficulty of reproducing and debugging flaky tests, the fixing step is rarely achieved by practitioners. 
    Future work should focus on providing tools that assist the root cause identification and reproduction of flaky tests.
    
\end{itemize}